\documentclass{eceasst}

\usepackage{wrapfig}
\usepackage{subfigure}
\usepackage{paralist}
\usepackage{soul}
\usepackage{setspace }
\usepackage{paralist, tabularx}
\usepackage[compact]{titlesec}         
\titlespacing{\section}{4pt}{2pt}{2pt} 
\AtBeginDocument{
  \setlength\abovedisplayskip{0pt}
  \setlength\belowdisplayskip{0pt}
  }

\title{VFSIE - Development and Testing Framework for  Federated Science Instruments } 
\author{Anees Al-Najjar\autref{1}\sponsor{}, Nageswara S. V. Rao\autref{1}\sponsor{}, Neena Imam\autref{1}\sponsor{},
Thomas Naughton\autref{1}\sponsor{}, Seth Hitefield\autref{1}\sponsor{}, Lawrence Sorrillo\autref{1}\sponsor{}, James Kohl\autref{1}\sponsor{}, Wael Elwasif\autref{1}\sponsor{},
Jean-Christophe Bilheux\autref{1}\sponsor{}, Hassina Bilheux\autref{1}\sponsor{}, Swen Boehm\autref{1}\sponsor{}, Jason Kincl \autref{1}\sponsor{} } 
\institute{\autlabel{1}  Oak Ridge National Laboratory, Oak Ridge, TN 37831 }

\abstract{Recent developments in softwarization of networked infrastructures combined with containerization of computing workflows promise unprecedented ``compute anywhere and everywhere" capabilities for federations of edge and remote computing systems and science instruments.
The development and testing of software stacks that implement these capabilities over physical  production federations, however, is not very practical nor cost-effective.
In response, we develop a digital twin of the physical infrastructure, called the Virtual Federated Science Instrument Environment (VFSIE). This framework emulates the federation using containers and hosts connected over an emulated network, and 
supports the development and testing of federation stacks and workflows.
We illustrate its use in a case study involving Jupyter Notebook computations and instrument control.} 

\begin{document}
\maketitle

\section{Introduction}
\label{sec:Introduction}

Experimental science workflows are often executed over a federation of science instruments and computing systems, including local edge systems and remote supercomputers, all connected over wide-area networks.
These workflows are complex with computations distributed across the federated computing platforms and also instruments that conduct experiments and collect measurements.
Recent developments in Software-Defined Infrastructures (SDI) promise the orchestration of federated resources and networks using software at unprecedented speed and scale.
Concurrently, the containerization of computations promises unprecedented ``compute anywhere and everywhere" capabilities across a continuum of computing platforms that may be located at the edge, cloud and remote supercomputing sites.
Furthermore, combined with notebook technologies, these computations can be effectively packaged with data and analytics modules as containers, and shipped and executed over various platforms.
 
\begin{figure*}[tb]
    \centering
    \setlength{\fboxsep}{3mm}
    \includegraphics[width=1\textwidth]{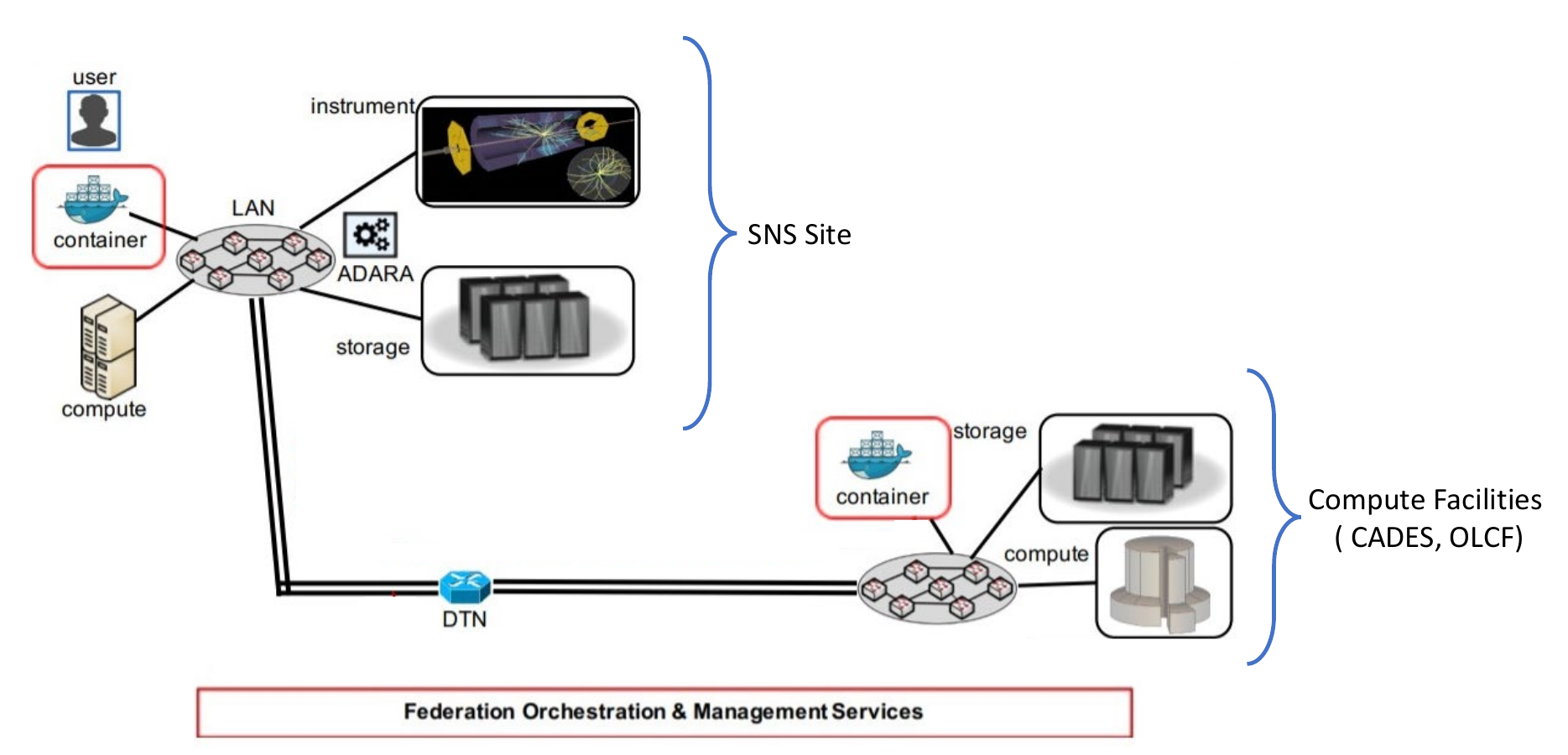}
    \caption[Compute mobility with SNS/HFIR]{SNS/HFIR workflows to leverage federated resources \cite{Naughtonetal2020smc}.}
    \label{fig:fedsnsmobility}
   \vspace*{-0.1in}
\end{figure*}

\section{Software Stack and Emulation Testbed}
\label{sec:Software_Stack_and_Emulation_Testbed}

A software framework and software stack, called the FedSci stack, are recently proposed by Naughton et al  \cite{Naughtonetal2020smc} to enable the science users to transparently execute workflows and the
facility providers  to contribute resources to participate in the federation.
This federation consists of experimental devices, such as beam-line instruments, computing servers and clusters, 
 and high-performance computing resources, including compute, network, storage systems.
An example use case involves a federation of
neutron imaging beam-lines (SNAP/BL-3, Imaging/CG-1D) at the Spallation Neutron
Source (SNS) and High Flux Isotope Reactor (HFIR) facilities and computing systems at Oak Ridge National Laboratory (ORNL), as shown in Figure~\ref{fig:fedsnsmobility}.
Here, custom codes generate tomographic reconstructions from x-ray images of a target, which are analyzed by science teams, often resulting in next target configurations to be used to collect measurements from additional vantage points. 
These codes are provided as Jupyter Notebooks.

The development and testing of the software stack for the federation and science workflow computations require the allocations of
significant resources and coordination among sites.
It is not very practical nor cost-effective, particularly during early functionality testing, to test them over the federation of production facilities.
Indeed, significant functionality testing and code debugging can be achieved without these physical facilities, thereby avoiding the use of valuable resources and mitigating disruptions due to unintentional code effects.
%
In response, we propose the Virtual Federated Science Instrument Environment (VFSIE) that emulates the federation using a combination of containers and hosts connected over an emulated network using ContainerNet. It is a digital twin that replicates the software environment and emulates the physical parts of federation.

VFSIE provides four distinct capabilities for science users, resource providers, federation maintainers and software stack developers:
\begin{compactitem}
\item [(i)] Science users execute their workflow codes in a manner identical to on physical infrastructure, namely, by building containers of their computations and launching them to hosts,  which are orchestrated by the FedSci stack.
\item [(ii)] Facility providers execute FedSci modules on emulated hosts or containers that represent their facilities. In particular, experimental facilities run Experimental Physics and Industrial Control System (EPICS) software, which is predominantly used by science facilities including SNS and HFIR.
\item [(iii)] Federation maintainers execute FedSci modules either as containers or as codes on designated hosts much the same way as over the physical infrastructure.
\item [(iv)] FedSci software developers execute modules on hosts or as containers for development and testing. They access docker containers from inside ContainerNet over the emulated network (as federation users) \textit{host-to-container connections} and  also via services and API's of host docker network, via \textit{container-to-container connections}; this flexible access makes it easier to test FedSci stack modules and science workflows. 
\end{compactitem}
VSFIE is executed inside a Virtual Machine (VM) that can be replicated  so that its components can be independently developed and tested, and uploaded to other sites. Upon maturity, these codes can be rolled into the physical infrastructure.


\section{Implementation and Use Case}
\label{sec:Implementation_and_Use_Case}

%
%

\begin{wrapfigure}{r}{0.5\textwidth}
\centering
\vspace*{-0.1in}
\includegraphics[trim=0cm 0cm 0cm 1cm, width=0.50\textwidth]{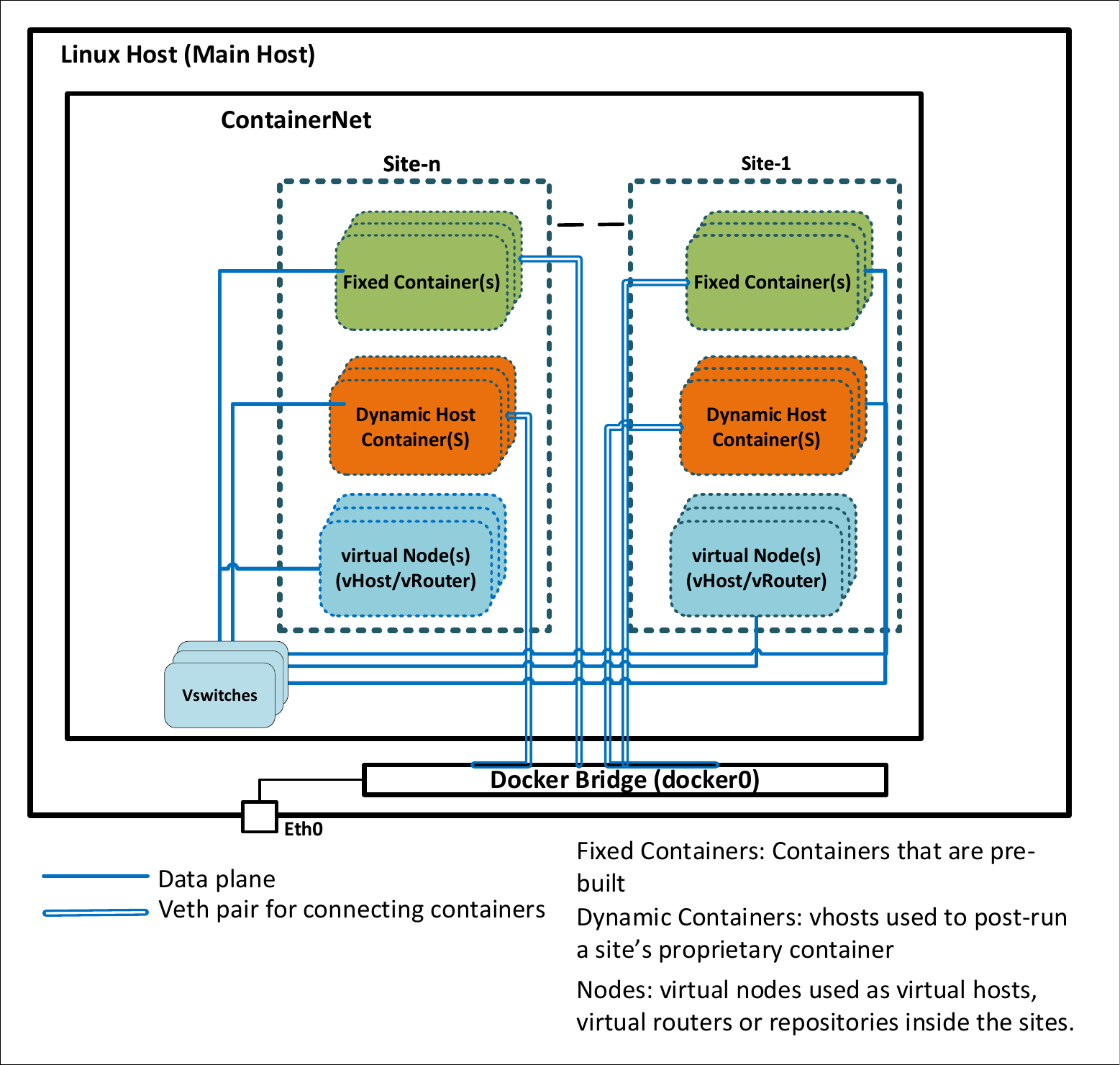}
\caption{VFSIE Architecture}
\label{fig:VFSIE_Architecture}
\vspace{-0.1in}
\end{wrapfigure}
Each federation site is emulated using a combination  of containers, hosts and its local-area network as shown in 
Figure~\ref{fig:VFSIE_Architecture}.
VFSIE allows for the execution of science workflow computations in the same way as by a science user on physical infrastructure. Science instruments are emulated as static containers associated with the sites.
The Virtual Science Network Environment (VSNE) \cite{Liuetal2018aiscience} supports the development of end-to-end network path provisioning using VMs and mininet to emulate sites, and VFSIE incorporates its basic design and networking components.

The emulation testbed consists of SNS site, and two separate ORNL computing facilities, 
Compute and Data Environment for Science (CADES) and Oak Ridge Leadership Computing Facility (OLCF), which are currently implemented as hosts and will be replaced by more accurate emulations.
SNS experimental facility is emulated using \textit{EPICS} software that controls beam-line targets as a fixed container, and the tomographic reconstruction code \textit{imars3d} is implemented as a container that can be ported and executed on CADES and OLCF hosts. The workflow containers are executed as native containers on hosts at three sites and FedSci stack modules are implemented as fixed containers at the sites.
The sites connect to gateway routers implemented as virtual hosts.

\begin{figure}[t!]
\subfigure[Science user workflow]{
\includegraphics[trim=0cm -2cm 0cm 0cm, width=0.45\textwidth,height=1in]{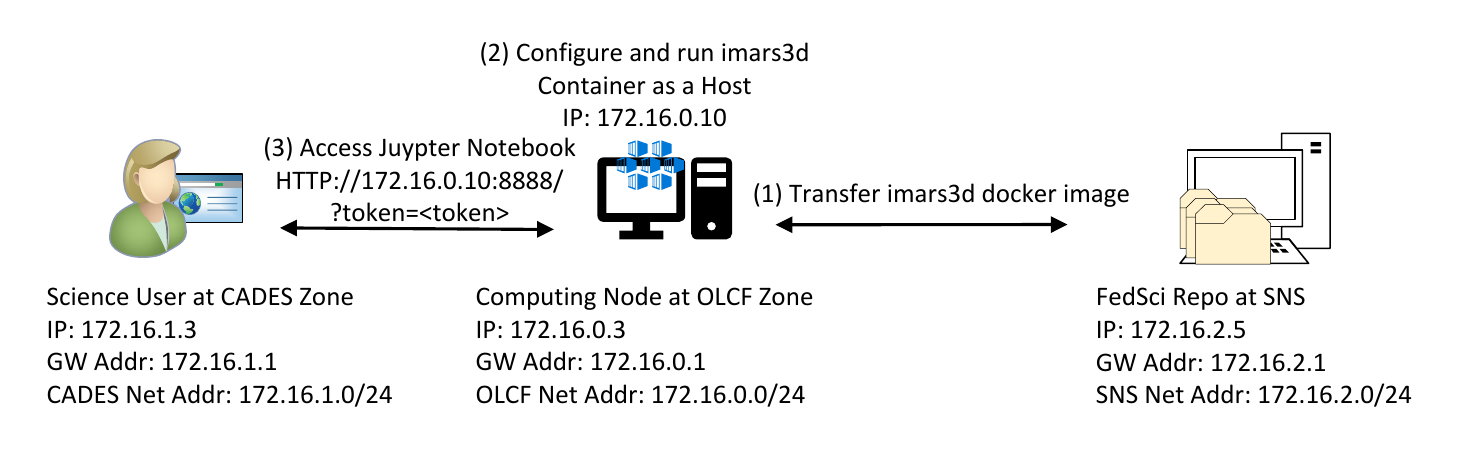}
\label{subfig:workflow}
}
\subfigure[Remote access to Juypter Notebook via browser]{
\includegraphics[width=0.45\textwidth,height=2in]{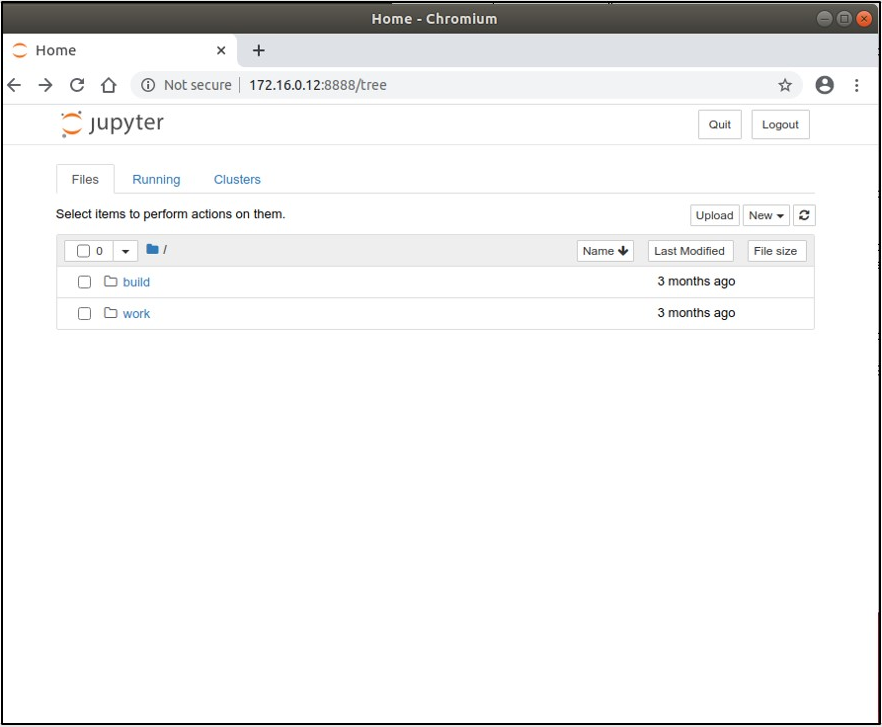}
\label{subfig:Requesting_access_to_Juypter_Notebook_from_the_Browser}
}
\caption{Workflow execution: science user accesses imars3d from CADES host.}
\label{fig:Accessing_imars3d_Juypter_Notebook_from_CADES}
\vspace*{-0.2in}
\end{figure}

An illustration of VFSIE functionality involves the execution of workflow tasks of a science user initiated from a CADES host in our use case, as depicted in Figure~\ref{subfig:workflow}.
FedSci host is setup with docker images for science user's workflow computations. The images are stored as \textit{tar} files using \textit{docker save} command. OLCF host is used to execute \textit{imars3d} code, and is accessed by the science user from CADES host.
The computing facilities and SNS site are in different networks. 
The imars3d container image is transferred to OLCF, where it is loaded using \textit{docker load} and run with mapping ports 8888:8888 by FedSci stack. A network bridge created at this node with its network interface and the virtual interface to allow imars3d container to be accessed via emulated network (outside host docker network). 
The science user accesses the Juypter Notebook of running imars3d container from CADES host via container IP address at OLCF host in its network (172.16.0.10/24).

Figure~\ref{fig:Accessing_imars3d_Juypter_Notebook_from_CADES} shows the orchestration of a science workflow that provides access to Juypter Notebook at OLCF from CADES host. The Chromium web browser is executed on CADES host with IP address 172.16.1.3. The access to the Juypter Notebook is then requested using container IP address and its external access port using \textbf{http://172.16.0.10:8888}. 
An access token is generated when \textit{imars3d} docker container is initiated, which is used by the science user to access data and services of Jupyter Notebook. Specifically, the browser loads the Juypter Notebook as a web service making it available to the science user at CADES host, as shown in  Figure~\ref{subfig:Requesting_access_to_Juypter_Notebook_from_the_Browser}.

\section{Conclusions and Future work}
\label{Conclusions}

VFSIE implementation in this paper demonstrates the feasibility of virtual frameworks for executing science workflows over emulated federated facilities.
The current VFSIE is under development and will be continually updated.
Our ongoing work involves the continued development of FedSci stack and analytics modules and their integration into VFSIE framework.
Future work may involve implementing more detailed emulations of sites and their multiple systems, and more complex federations with multiple facility sites distributed across wide-are networks.

\begin{acknowledge}
This work is funded by SDN-SF and RAMSES projects, U.S. Department of Energy, and Laboratory Directed R\&D  project,  and performed at Oak Ridge National Laboratory managed by UT-Battelle, LLC under Contract No. DE-AC05-00OR22725.
\end{acknowledge}



\begin{spacing}{0.5}
\bibliographystyle{eceasst}
\bibliography{nwref}
\end{spacing}
\end{document}